# Quantum Mechanical Relations for the Energy, Momentum and Velocity of Single Photons in Dispersive Media

by

Robert J. Buenker

Bergische Universität-Gesamthochschule Wuppertal, Fachbereich 9-Theoretische Chemie, Gaussstr. 20, D-42097 Wuppertal, Germany

and

Pedro L.  Muiño

Department of Chemistry, Mathematics, and Physical Science, St. Francis College, Loretto, Pennsylvania 15940 USA

**Abstract**

Attempts to explain the refraction of light in dispersive media in terms of a photon or "corpuscular" model have heretofore been unable to account for the observed decrease in the speed of light as it passes from air into a region of higher refractive index n such as water or glass. In the present work it is argued on the basis of the quantum mechanical relations $p = \hbar k$ and $E = \hbar \omega$ that the energy of photons satisfies the equation $E = pc/n$. It is possible to obtain an exact prediction of the observed speed of the photons in a given medium by application of Hamilton's equations of motion to the above formula, but at the same time to conclude, in agreement with the arguments of Newton and other classical physicists, that the photon momentum increases in direct proportion to n, thereby producing the well-known bending of light rays toward the normal when entering water from air. The corresponding relativistic particle theory of light indicates that the potential V encountered by the photons in a given medium is attractive for n > 1 and is *momentum-dependent*, which suggests the microscopic interactions responsible for the refraction of light are non-Coulombic in nature and are instead akin to the spin-orbit and orbit-orbit terms in the Breit-Pauli Hamiltonian for electrons moving in an external field. The present theory concludes that the reason photons are slowed down upon entering water from air is that their relativistic mass p/v increases faster with n than does their momentum, which in turn requires that Einstein's famous $E = mc^2$ formula does not hold for light dispersion because the energy of the photons is expected to be the same in both media. In summary, all the known experimental data regarding light dispersion can be successfully explained in terms of a particle theory of light once it is realized that photons possess exceptional properties by virtue of their zero proper mass and their capacity to undergo electromagnetic interactions with surrounding media.

## I. Introduction

The refraction of light as it passes through transparent media is a phenomenon of everyday experience which has occupied the attention of scientists dating back at least to the time of the ancient Greek philosophers. Since the experiments of Foucault in 1850, it has been known that light moves more slowly in water and other materials than it does in free space. Newton [1] and his followers had come to the opposite conclusion based on his corpuscular theory of light, and this fact had a decisive effect in promoting the competing wave model of electromagnetic radiation [2, 3]. It was simply argued that if light were really composed of particles, classical mechanics should have been able to predict that it would be slowed down in dispersive materials, and having failed in this, the above premise should be discarded entirely. The pioneering developments of the early 20th century [4, 5] caused a rethinking of this conclusion, however, so that there is now a consensus among physicists that some experiments are best explained in terms of light waves, while others seemingly require a particle formulation to remain consistent with the observed results. Moreover, the concept of wave-particle duality has been generalized by de Broglie [6] to apply to all types of matter.

The question that will be explored in the present study is whether there is really no way of understanding light dispersion in terms of a particle model. Recent experiments [7] have shown, for example, that single photons travel through glass at the group velocity of light $v_g$ = $c/n_g$. The technique employed makes use of a two-photon interferometer and thus takes advantage of the wave properties of light, but one can reasonably conclude that what is being measured is simply the speed of individual photons in the apparatus. To explore this point further, new experiments have been carried out in our laboratory [8] in which time-correlated single-photon counting (TCSPC) detection has been employed to measure the ratio of the speeds of light in air and water. The most interesting aspect of this study was that the shapes of the photon counting profiles (instrument response) are not affected by the medium through which the light passes (over a distance of nearly 1.0 m). It is possible to explain this result in a very simple manner, namely that, as in the earlier experiments of Steinberg et al [7], *each* photon is decelerated by an amount which is inversely proportional to the group index of refraction $n_g$ as it passes from air into a dispersive medium. These findings certainly add support to the particle theory of light advocated by Newton [1], but at the same time underscore the need to better understand why the classical theory leads to an erroneous prediction of the dependence of the speed of light on the nature of the medium through which it passes.

## II. Photon Momentum in Dispersive Media

Newton's argument was based on observations of the refraction of light in dispersive media (see Fig. 1). According to Snell's Law of Refraction, the angles of incidence $\theta_1$ and refraction $\theta_2$ at an interface are inversely proportional to the respective indices of refraction which are characteristic for each medium:

$$n_1 \sin \theta_1 = n_2 \sin \theta_2. \tag{1}$$

The bending of light rays was regarded as evidence for the existence of a force acting at the interface between two media. Since light always travels in a straight line within a given medium, it follows that the corresponding potentials are constant throughout and the resultant force acting on the light must be in the direction normal to the interface. Because of Newton's Second Law this means that the component of the momentum of the particles of light which is parallel to the interface must also be constant; hence, according to Fig. 1,

$$p_1 \sin \theta_1 = p_2 \sin \theta_2, \tag{2}$$

which implies by comparison with eq. (1) that the total momentum of the particles p is always proportional to the refractive index n of a given medium. Since light is bent more toward the normal in water than in air, one is led unequivocally to the conclusion that the momentum of the photons is greater in water.

From there it was only a short step for Newton to conclude that the velocity of light v must also be greater than in air, since by definition,

$$p = mv, \tag{3}$$

and there was no evidence at that time to indicate that the inertial mass m of mechanical particles could be anything but a constant. This conclusion becomes far less obvious, however, once the results of Einstein's Special Theory of Relativity [9] are taken into consideration. It is known that the mass of a particle varies with the relative velocity of the observer, for example, and also that mass is not conserved in reactive processes.

On closer examination, it is clear that Newton's arguments are only directly applicable to the *momentum* of photons in dispersive media. Since there have never been any quantitative measurements of photon momenta in condensed media, such as by the observation of x-ray scattering with electrons therein or of nuclear recoil following high-energy emission processes, it seems fair to say that there is still a good possibility that eq. (2) is correct. Indeed, there is independent evidence obtained from application of the quantum mechanical relation,

$$p = \hbar \, k, \tag{4}$$

to the wave theory of light that this is so ($k = 2 \pi / \lambda$). The fundamental equation for the phase velocity of light in dispersive media is:

$$\omega / k = c / n. \tag{5}$$

In this equation it is known that the frequency $\omega$ is independent of n and therefore that the wave vector k is proportional to n. Comparison with eqs. (2) and (3) shows that according to Newton's theory, the momentum of the photons must be proportional to k, consistent with eq. (4).

The latter equation is acknowledged to be valid for photons in free space [10] and, following de Broglie's hypothesis [6], to hold for free particles in general. Furthermore, in the Davisson and Germer experiments [11], it was necessary to take into account the fact that the momentum of the electrons increases upon entering the Ni crystal interior in order to correctly predict the wavelength of the maximum in the observed electron diffraction pattern. This shows that eq. (4) is also valid for particles in the presence of a potential. Dicke and Wittke [12] have also pointed out that a consistent definition of a refractive index can be made for electrons passing between different regions of constant potential if one assumes the de Broglie relation, in which case one again finds that p is proportional to n. It is interesting to note that it is not necessary to know the value of $\hbar$ in eq. (4) to arrive at such relationships, only that there is a proportionality. Thus, the pioneering experiments of the late 19th century that led to the formulation of the quantum theory of matter are not actually needed to infer that the momentum of photons in dispersive media is directly proportional to the wave vector k, and is thus larger in water and glass than in air. Since both the photon energy E and the light frequency ω are independent of n, it is also possible to infer their proportionality in light refraction without knowledge of the Planck equation [13],

$$E = \hbar\,\omega. \tag{6}$$

This relationship can be combined with eq. (4) to form a single relativistically invariant expression for the corresponding four-vectors [10], although this could not be deduced from what was known to Newton because it requires that the proportionality factor be the *same* in both equations.

The wide range of applicability of eqs. (4,6) therefore speaks strongly for Newton's conclusion that the momentum of the elemental particles of light increases as they enter a medium of higher n. It is interesting to compare this with the interpretation based on the wave theory of light where one traditionally speaks of marching soldiers entering into a muddy field and decreasing their spacing (wavelength) while maintaining their original cadence (frequency) [14]. In order to avoid breaking ranks it is argued that a change in direction is required which exactly corresponds to what occurs when light is refracted, namely bending toward the normal and a decrease in phase velocity. Clearly, in this view work is being done by the photons as they enter the medium of higher n, and their individual momenta therefore *decrease* as the wavelength is narrowed. While this explanation is consistent with all the known experimental data regarding the refraction of light, it is seen to stand in disagreement with the quantum mechanical relation of eq. (4). Newton's version of the theory at least does not have this problem, and yet it also explains the bending of light in a consistent manner. It implies that the photon momentum satisfies eq. (2) and thus is proportional to n, rather than decreasing with it as in the competing model.

The quantum mechanical relations of eqs. (4,6) also lead to a definite prediction about the nature of the potential V for light in a dispersive medium. According to special relativity,

$$E - V = (p^2 c^2 + \mu^2 c^4)^{1/2},$$ (7)

where $\mu$ is the rest mass of the particle. In the case of light, $\mu = 0$ and thus in this special case,

$$E - V = pc.$$ (8)

Applying eqs. (4,6) to the experimental relation of eq. (5) gives

$$E = pc / n,$$ (9)

which , when combined with the previous equation, yields

$$V = (1-n) E = (1-n)\ \hbar\, \omega.$$ (10)

According to this result, the potential V is attractive for media such as water and glass with n > 1, which is clearly consistent with the classical physicist's view that the momentum of the particles of light increases in them relative to air or free space, but which is incompatible with the model of a muddy field in the corresponding wave theory. The fact that V increases with the energy of the photons according to eq. (10) indicates that it is not an electrostatic (Coulomb) potential, but this is not surprising in view of the electrical neutrality of photons. Evidently, the potentials in question are *momentum-dependent*, which suggests that they are akin to interactions such as spin-orbit and orbit-orbit coupling which are met in relativistic quantum mechanical formulations of the interactions of electrons in atoms and molecules [15]. This observation is thus also consistent with the fact that light is a distinctly relativistic phenomenon. According to eqs. (1,2), the amount by which the light rays change direction when entering a given medium, as quantified by the value of the index of refraction n, is a direct measure of the relative gain (loss) in momentum which the photons experience in the process.

### III. Calculation of the Photon Velocity

The preceding discussion emphasizes that any attempt to explain the above experimental results in terms of a particle model of light must come to grips with a simple fact. Classical mechanics demands that the momentum of the photons be greater in water than in air, even though the measurements show that the opposite relationship holds for their velocity. If one assumes, as Newton did, that the mass of the particles of light is the same in all dispersive media, one is forced to conclude from the definition of eq. (3) that v = nc for the photon velocity. As mentioned in the Introduction, this incorrect result was taken as proof that light consists of waves rather than particles, but it is clear from the TCSPC experiments [8] that single photons with a well-defined momentum and energy are being detected one at a time at a photomultiplier tube, *all* with a smaller velocity in water than in air.

There is a simple way to avoid this dilemma, however, and that is to compute the photon velocity using Hamilton's canonical equations of motion [16]:

$$v = dE / dp. \qquad (11)$$

Applying this with the aid of eq. (9) leads to the observed result [17],

$$v = c / n + p \, d(c/n)/dp = c / n + k \, d(c/n)/dk, \qquad (12)$$

where eq. (4) has been used to eliminate p on the right-hand side. The same result is obtained by using both eqs. (4,6) directly, which gives

$$v = dE/dp = d\omega / dk = c / n_g = v_g, \qquad (13)$$

the group velocity of light. Either way, it is clear that one needs the benefit of quantum mechanics to obtain the above result. It is therefore easily understandable why Newton was not able to to deduce the correct value for the light velocity, but at the same time it needs to be emphasized that this failure in no way invalidates his prediction of an increase in the momentum of the light particles as they enter a medium of higher index of refraction.

To obtain the same result for the velocity in the wave theory, one must explain why light travels with the group velocity rather than with the phase velocity of eq. (5), $v_p = c / n$. It is necessary to assume that light waves with a definite frequency and wavelength are no longer monochromatic when they enter a dispersive medium [17,18]. By analogy to Rayleigh's theory of sound [19], the superposition of two such wave functions can be written as the product

$$\Psi = A\cos (\omega \, t - k \, x) \cos (\Delta \, \omega \, t - \Delta \, k \, x), \qquad (14)$$

where $\Delta\omega$ and $\Delta k$ are the respective changes in the frequency $\omega$ and wave vector $k$ relative to their unperturbed values in free space (A is the constant amplitude).

It should be noted that all such arguments are purely theoretical, however, as no one has ever been able to measure either $\Delta\omega$ or $\Delta k$. This is in contrast to the case for sound or ocean waves. If two musical instruments are out of tune, one hears both the main frequency $\omega$ and the beat frequency $\Delta\omega$. Similarly, if one drops a rock in a pond it is possible to observe both the envelope that corresponds to the average amplitude variation moving with the group velocity and the individual wavelets moving within them at a different (phase) velocity. In the case of light traveling in dispersive media there is no doubt that waves of a given frequency $\omega$ and wavelength k exist, but nothing is ever found to move with the corresponding (phase) velocity. The Fizeau method for measuring the speed of light involves the interference of waves with the main (carrier) $\omega$ and k values [18,20], but the actual light speed which is measured is never the ratio of these two quantities ($v_p$) but rather the group velocity $d\omega / dk$. To explain this fact, one argues that the requisite $\Delta\omega$ and $\Delta k$ values of the wave groups or packets are simply too small to be observed. If this is the case, however, it means that the period and the wavelength of the amplitude variations must be extemely large, that is, barely noticeable over any finite time of measurement. Under the circumstances one can reasonably ask why the waves only travel at the group velocity when neither the elapsed time nor the distance traveled during the measurement is ever large enough to allow for observation of the frequency or wavelength of the corresponding periodic motion.

A similar problem does not exist in the particle model of light because the derivative $dE / dp$ in Hamilton's canonical equations [16] implies nothing about the distribution of photon velocities. One simply knows that v can be computed from eq. (11) with knowledge of the variation of E(p) in the immediate neighborhood of the particle's actual momentum. This numerical procedure is valid even if all the particles are moving at exactly the same speed, which would correspond to perfectly monochromatic waves in the competing model. The wave theory has no other means of explaining why the $d\omega / dk$ derivative is measured for the light velocity than to assert that there is a secondary dispersion effect which prevents waves from retaining their initial frequencies and wavelengths even though the resulting variations in these quantities are far too small to ever be measured. In summary, the fact that the speed of light in dispersive media is always found to be equal to the group velocity does not at all prove that the elemental constitution of light is other than particles. Instead, it adds support to the conclusion that Hamilton's canonical equations of motion are applicable in this case, and that the way to compute the derivative in eq. (11) is to infer the variation of the photon's energy with momentum from the quantum mechanical wave-particle relations of eqs. (4,6).

**IV. Photon Mass**

In the preceding discussion the mass of the photon has been mentioned several times without coming to any conclusion about what it is or, for that matter, whether it makes sense to even talk about photon mass. We refer to the relativistc mass m, rather than the rest or proper mass $\mu$, which has been assumed to be zero throughout. In special relativity $m = \gamma\mu$, with $\gamma = (1-v^2 / c^2)^{-1/2}$, and so for light traveling with $v = c$ one cannot obtain a definite result from this formula since $\gamma = \infty$. This leaves open the possibility that $m \neq 0$ in this case, however. Indeed, it has been noted [21] that one can consistently define the mass of photons in free space as $E / c^2 = \hbar \, \omega / c^2$. This even works in explaining the gravitational red shift of light [22].

In each case above reference is made to photons moving in free space, however, so the question that remains is what can be said about their mass in dispersive media. For this purpose it is helpful to return to the definition of (inertial) mass in Newton's famous $F = ma$ relation, or in its time-integrated form, eq. (3). If the arguments in Sect. II about the momentum of photons being proportional to the refractive index of the medium are correct, it follows from eqs. (6,9) that

$$p = n \hbar \, \omega / c. \qquad (15)$$

Both measurements and theory find that the corresponding velocity is $v = c / n_g$, with $n_g$ defined in eqs. (12,13). Combining these two results in eq. (3) gives

$$m = nn_g \hbar \, \omega / c^2, \qquad (16)$$

which implies that the relativistic mass of the photon increases even faster with n under normal dispersion than does the corresponding momentum, hence producing the observed decrease in light velocity v.

This solution has some interesting consequences, however. Since $E = \hbar \omega$ is independent of n, it means that

$$E = m\, c^2 / n\, n_g, \qquad (17)$$

i.e., the famous $E = mc^2$ energy/mass equivalence relation [9] breaks down in the case of photons in dispersive media. If this were not the case, $E = \hbar \omega = mc^2$ would hold because of eq. (6). The photon mass would thus be independent of the dispersive medium, just as Newton assumed, and hence, there would be no recourse in the particle model of light but to assert that *both* the momentum and velocity of the photons are greater in water than they are in air. Eq. (16) gives a clear alternative to such a conclusion by allowing the photon mass to vary in going from one dispersive medium to another while continuing to insist that the photon energy, $\hbar \omega$, is a constant of motion. It should also be noted that the formula normally used for the velocity of elementary particles [23],

$$v = pc^2/E, \qquad (18)$$

assumes that $E = mc^2$ is valid, thereby reducing to the definition of eq. (3). Substitution of the E(p) relation of eq. (9) thus again leads to an incorrect prediction of $v = nc$. Another seemingly proper means of computing the mass of particles in the presence of a potential V consists of equating the quantity on the right-hand side of eq. (7) with $mc^2$ [24]. Combining this with eq. (8) for the special case of photons in dispersive media again leads to an incorrect result, however, because it would mean that $v = p / m = c$, independent of n.

The point that needs to be emphasized is that the failure of these various assumptions to explain the observed variation in the speed of light in dispersive media does not prove that it is unreasonable to employ a particle model of light in this application. A clear alternative exists, namely that photons are exceptional particles which have an unusual variation of energy with momentum and relativistic mass, as made explicit in eqs. (9) and (17), respectively.

A general expression for the relativistic mass of particles moving under the influence of a potential which is in agreement with all known experimental data, including the case of light in dispersive media, can be obtained from Hamilton's canonical equations, however:

$$m = p \, dp \, / \, dE = \frac{1}{2} \, \frac{d(p^2)}{dE} \, . \qquad (19)$$

For free particles $E = p^2 / 2m$ in the nonrelativistic regime and one merely sees that the above definition is consistent, whereas for much higher speeds, $E = (p^2c^2 - \mu \, c^2)^{-1/2}$, and one finds that $m = E / c^2$ by assuming that $\mu$ is constant. For photons in a dispersive medium of refractive index n, the only means at present of evaluating eq. (19) is by assuming the quantum mechanical relations of eqs. (4,6), or more specifically the E(p) relation of eq.(9), in which case eq. (16) results. An *ab initio* calculation of the refractive index n is quite difficult, primarily because the condensed phase systems with which the photon is interacting are exceedingly complex. For all known particles of nonzero proper mass (minimally in the MeV range), typical potentials are presumably too weak (on the order of tens of eV) to expect any measurable deviations in their relativistic mass relative to the conventional $E / c^2$ formula.

**V. Conclusion**

Recent measurements of the speed of single photons in dispersive media have provided a stimulus to reexamine the particle theory of light. The argument that only a wave theory can successfully account for the refraction of light at the interface between two media has been shown to overlook several important possibilities. Application of the quantum mechanical results of eqs. (4) and (6) to the experimental relation for the phase velocity of light indicates that the photon energy satisfies the equation, $E = pc/n$, which in turn leads one to conclude that the momentum of the photons is directly proportional to the refractive index in a given medium. At the same time, differentiation of the same equation with respect to p leads to the observed relationship, $v = c/n_g$. The surprising result of this new version of the particle theory is thus that when light passes into a region of higher n, the speed of each photon should decrease while the corresponding momentum increases.

The original mechanical theory proposed by Newton showed first and foremost that the momentum of the light corpuscles should increase in a region of higher n, and this argument is simply verified by the quantum mechanical relations. There is apparently no experimental evidence to contradict this position, since standard methods of determining the momentum of photons, such as the Compton effect or nuclear recoil in radiative emission, have only been carried out under near vacuum conditions. The fact that Newton then concluded that the speed of light should also increase in a medium of higher n followed from his belief that the inertial mass of the light corpuscles should remain constant. By combining the above theoretical result for the dependence of photon momentum on refractive index with experimental results for the speed of light in dispersive media, one is led to a quite different conclusion, however. Accordingly, the photon mass is proportional to the product of both n and $n_g$, and since it increases faster when the photons pass from air to water than does the corresponding momentum, a decrease in the speed of light is observed. It should be clear in this discussion that the rest mass $\mu$ of the photons is assumed to be zero throughout and thus is to be clearly distinguished from the relativistic mass m of eq. (3). It should also be emphasized that the proposed variation of the photon mass with n and $n_g$ of the medium necessarily means that Einstein's $E = mc^2$ relation does not hold for photons in the presence of dispersive forces, since there is no alternative but to assume that the photon energy does not change from one medium to another.

In summary, it is not true that experimental measurements of the speed of light in water and other transparent media prove that a particle or "corpuscular" theory is incapable of accounting for the effects of dispersive forces. Indeed, the most straightforward means of explaining the results of the recent TCSPC [8] experiments as well as the light speed measurements of Steinberg et al. [7] is to assume that single photons corresponding to a given wavelength of light are all uniformly decelerated when entering a medium of larger $n_g$. The only way to test this theory further is to carry out new experiments which allow one to measure the momentum of the individual photons rather than just their speed. The quantum mechanical relation, p = $\hbar$ k, gives a strong indication that photon momenta do increase with n, in agreement with at least that part of Newton's original theory. Given the fact that his incorrect prediction about the speed of light in dispersive media can be easily rectified with the aid of quantum mechanics, there is good reason to believe that the main thrust of his arguments, namely that light is composed of particles which we now call photons, still remains a viable option for the description of optical phenomena.

**Acknowledgment**

This work was supported in part by the Deutsche Forschungsgemeinschaft within the Schwerpunkt Programm *Theorie relativistischer Effekte in der Chemie und Physik schwerer Elemente*. The financial support of the Fonds der Chemischen Industrie is also hereby gratefully acknowledged.

Figure Captions

Fig. 1. Schematic diagram showing the refraction of light at an interface between air and water.  The fact that the light is always bent more toward the normal in water (Snell's Law) led Newton to believe that there is an attractive potential in the denser medium which causes the particles of light to be accelerated.

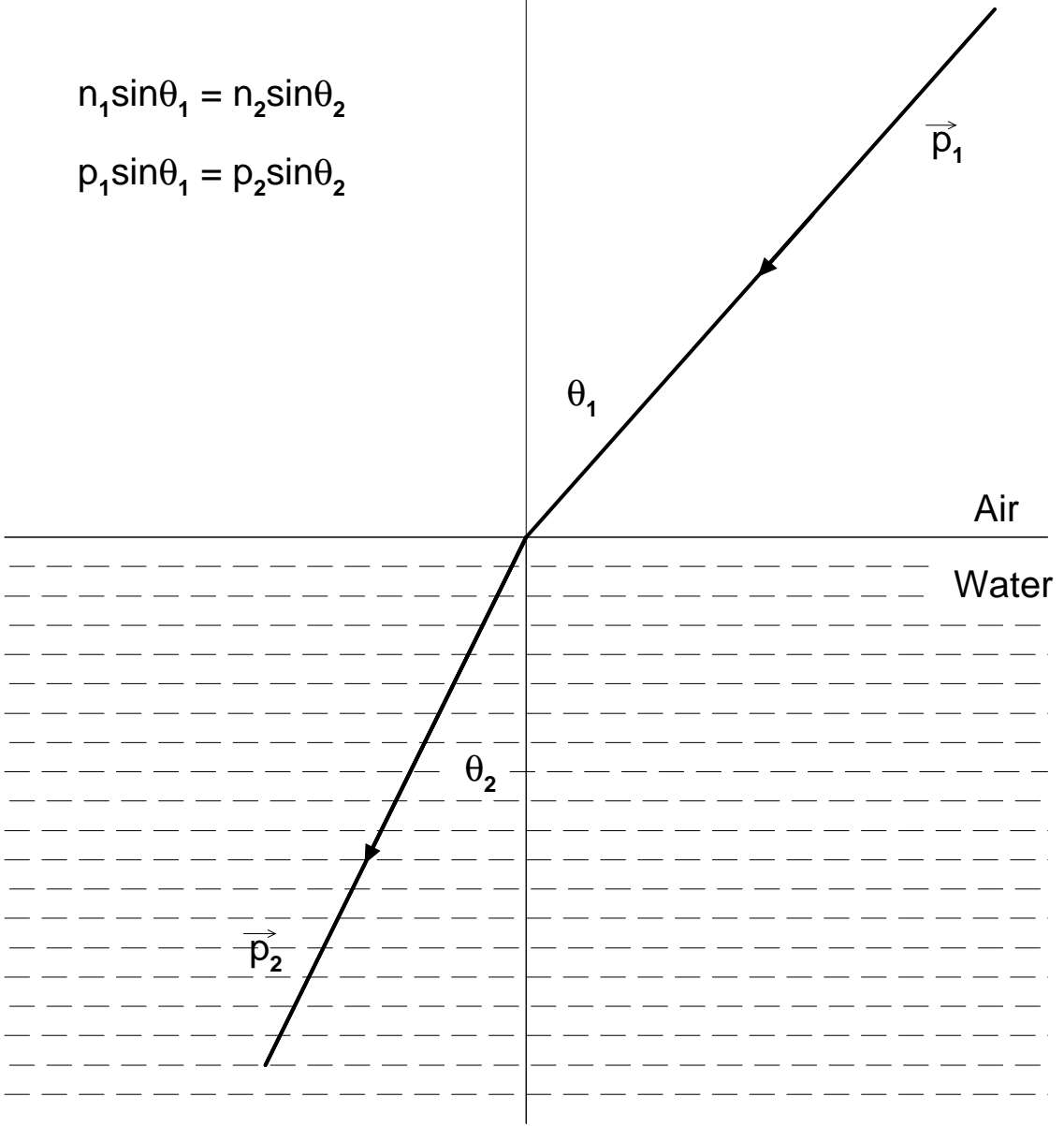